\documentclass{article}
\usepackage{graphicx}%
\usepackage{multirow}%
\usepackage{amsmath,amssymb,amsfonts}%
\usepackage{amsthm}%
\usepackage{mathrsfs}%
\usepackage[title]{appendix}%
\usepackage[dvipsnames]{xcolor}
\usepackage{textcomp}%
\usepackage{manyfoot}%
\usepackage{booktabs}%
\usepackage{algorithm}%
\usepackage{algorithmicx}%
\usepackage{algpseudocode}%
\usepackage{listings}%
\usepackage{gensymb}
\usepackage{placeins}
\usepackage{authblk}
\usepackage{gensymb}
\usepackage{hyperref}

\DeclareMathOperator\erf{erf}
\DeclareMathOperator\Erf{Erf}

\newcommand\EJsout{\bgroup\markoverwith{\textcolor{RedOrange}{\rule[0.5ex]{2pt}{1pt}}}\ULon}

\newcommand{\ssSec}[2]{\subsubsection{#1 \label{Sec:#2}}}

\newcommand{\pare}[1]{\left(\, #1 \, \right)}

\newcommand{\bra}[1]{\left[\, #1 \, \right]}

\newcommand{\cG}{{\mathcal G}}

\newcommand{\cK}{{\mathcal K}}

\newcommand{\cS}{{\mathcal S}}

\newcommand{\cP}{{\mathcal P}}
\newcommand{\cR}{{\mathcal R}}

\begin{document}

\title{The refresh rate of overhead projectors may affect the perception of fast moving objects: a modelling study}

\date{}

\author[1]{J\'er\^ome Emonet}
\author[1]{Bruno Cessac}
 
\affil[1]{Université Côte d’Azur, Inria Biovision Team and Neuromod Institute}

\maketitle

\begin{abstract}
Using simulation and a simple mathematical argument we argue that the refresh rate of overhead projectors, used in experiments on the visual system, may impact the perception of fast moving objects at the retinal and cortical level (V1), and thereby at the level of psychophysics.
\end{abstract}

%

\section{Introduction}\label{Sec:Introduction}

To explore our visual environment, our eyes constantly make saccades. This fast and unconscious movement can reach a maximal speed of $200\degree/s$ \cite{westheimer:54} that we do not perceive though, because it is masked. This phenomenon, known as saccadic omission \cite{campbell-wurtz:78}, allows us to remain unconscious of the rapid movements caused by our own eyes, thereby maintaining a stable representation of our environment. To better understand this phenomenon, psychophysicists simulate saccades by the rapid movement ($200$ $\degree/s$) of a bar in the visual field of individuals maintaining a fixed gaze \cite{duyck-wexler-etal:18}. This type of stimulus is usually displayed 
by an overhead projector with a refresh rate of $60$ Hz. However, when considering an object moving at $200$ $\degree/s$, i.e. covering $3.33$ degree of visual angle at each frame, the effective motion provided by this projector is  quite "jerky" and discontinuous. It is then rather natural to ask whether this lack of smoothness is perceived by the subjects. In this context, Mark Wexler, from the Integrative Neuroscience and Cognition center,  (unpublished work), had the idea to 
confront the perception of simulated saccades when the frame rate of the overhead projector is changed from low ($60 $ Hz) to high ($1440$ Hz). He highlighted a clear difference in the perception of subjects. At high frame rate, the bar is perceived as more distinct, with an apparent lower speed, a shorter amplitude of motion and without the smear that was perceived at low frame rate. 
This leads one to question how the frame rate impacts the perception of a subject. What makes our visual system able to discriminate between a high and low frame rate stimulus? At what level of the visual system could this discrimination take place? 

These questions are certainly quite difficult and maybe impossible to answer from a neuro-physiological perspective in the current state of the art. More humbly, as modellers, we would like to point out in this short communication how a simple mechanism of linear integration, taking  already place at the level of the early visual system (retina and V1) can display a response to a fast moving object baring some analogy with what was observed by M. Wexler.
For this, we use a retino-cortical (V1) model that we initially designed to study anticipation in the early visual system \cite{emonet-souihel-etal:24} and briefly introduced in the section \ref{Sec:Model}. We carry out simulations featuring the retina and V1 response to a fast bar moving at $200 \degree/s$  projected to the retina with a high ($1440$ Hz) or low ($60$ Hz) frame rate (section \ref{Sec:Impact_frame_rate}). In this context, the difference in response that we observe can be easily explained  using linear response as we discuss in a short mathematical analysis presented in section \ref{Sec:params}.

\section{Model}\label{Sec:Model}

\begin{figure}
\centering
\includegraphics[width=0.85\textwidth]{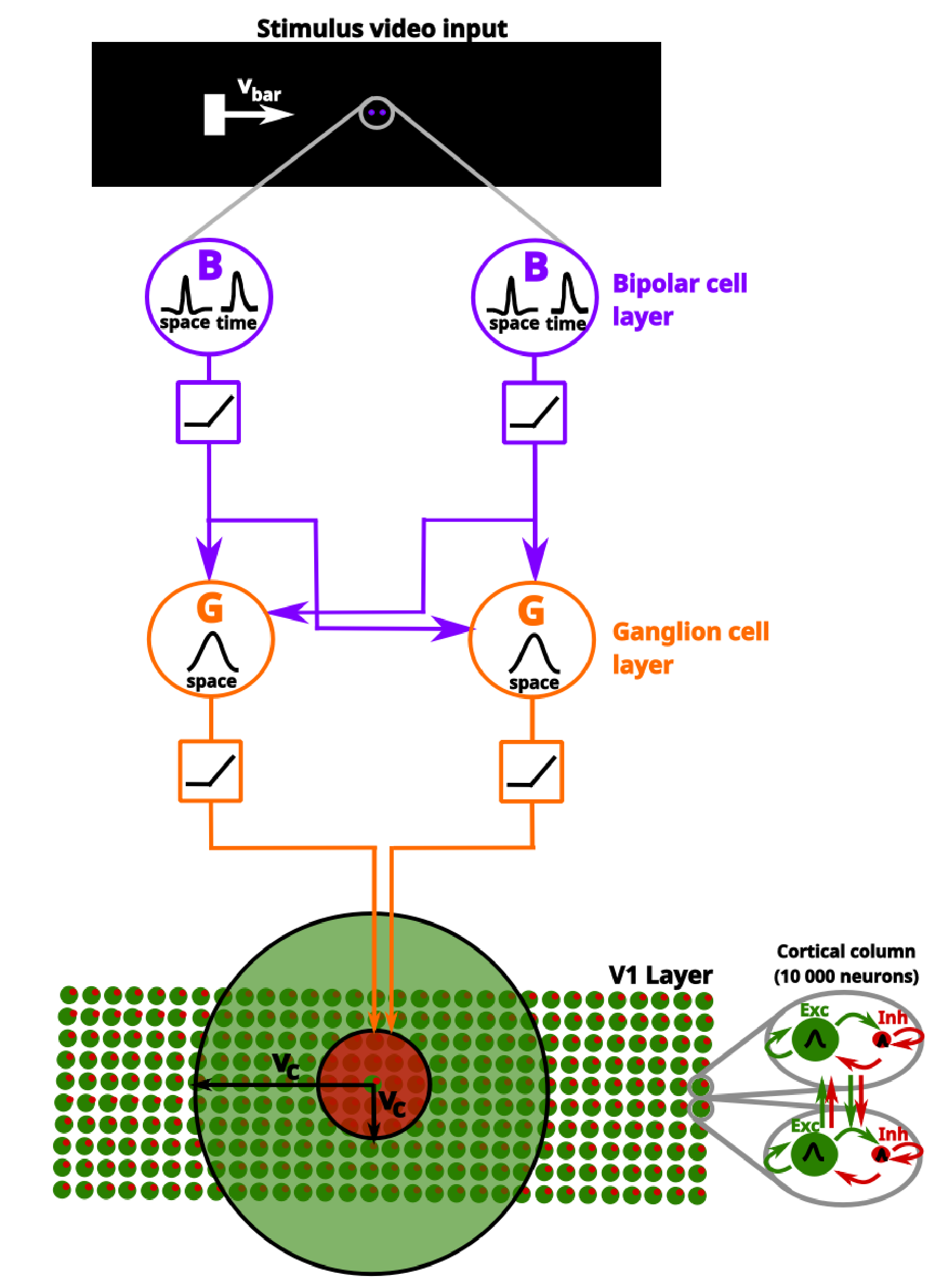}
\caption{\textbf{Synthetic view of the retino-cortical model.} A stimulus is perceived by the retina, triggering a response. \textbf{From top to bottom}: The stimulus is first convolved with a spatio-temporal receptive field (black traces labelled "space" and "time" in the purple circles) that mimics the coordinated activity of photoreceptors and horizontal cells in the outer plexiform layer (OPL), and passed on to bipolar cells (purple circles). This response is corrected by a low-voltage threshold (purple squares). The responses of the bipolar cells (BCs) are then pooled to the retinal ganglion cells (RGCs, orange). The firing rate of an RGC is a sigmoidal function of the voltage (orange square). The RGC response (firing rate) is sent to cortical columns in the primary visual cortex (V1), shown at bottom right as two interconnected mean field units (small circles), corresponding to the excitatory (green) and inhibitory (red) populations, respectively. A sketch of this connection is shown in the right zoom (grey ellipses). Cortical columns are connected by an excitatory (large green circle) and inhibitory (large red circle) horizontal connection. Note that we assume the same conduction velocity $v_{c}$ for both connections.} 
\label{Fig:Schema_modele_retino_cortical_simplified}
\end{figure}

The reader can refer to the paper \cite{emonet-souihel-etal:24} for more detail on the model. A synthetic description is given in Fig. \ref{Fig:Schema_modele_retino_cortical_simplified}.

The retinal model was inspired by the papers \cite{berry-brivanlou-etal:99,chen-marre-etal:13} and developed within our team in previous works \cite{souihel-cessac:21,cessac:22,kartsaki-hilgen-etal:24}. In its most general version it is composed of a hierarchy of retina sub-layers, bipolar cells, amacrine cells, retinal ganglion cells, connected together to mimic the real retina connectivity and containing gain control and rectification.
Here, we simplified it so that it is only composed of two bidimensional layers of bipolar (BC) and ganglion (RGC) cells (Fig. \ref{Fig:Schema_modele_retino_cortical_simplified}) where BCS are connected to RGCs by a Gaussian pooling, following \cite{berry-brivanlou-etal:99, chen-marre-etal:13}. BCs receive an input which is the convolution of the visual stimulus (here a movie) with a spatio-temporal kernel (see eq. \eqref{eq:Vdrive} below). BCs integrate this response via a rectified linear integrator and excite the RGCs. 

The retinal model is connected to a cortical model (the thalamus is transparent here, just considered as a relay) consisting of a grid of mean fields cortical columns based on Adapting Exponential (AdEx) neurons \cite{zerlaut-chemla-etal:18, di-volo-romagnoni-etal:19}. Cortical columns represent the spatial average of cortical neurons at a space scale roughly corresponding to one pixel of voltage sensitive dye imaging (VSDI signal) \cite{chemla:10}. Each modelled cortical column is composed of an excitatory and an inhibitory population of neurons. The response of cortical columns is represented by their simulated VSDI signal calculated from the average voltage of the two populations cortical populations. The model parameters have been tuned according to the paper
\cite{emonet-souihel-etal:24}. The simulations have been done with the Macular software (\url{https://team.inria.fr/biovision/macular-software/}) created at INRIA Sophia-Antipolis.

\section{Results}\label{Sec:Results}

\subsection{Impact of the frame rate}\label{Sec:Impact_frame_rate}

\begin{figure}[!ht]
\centering
\includegraphics[width=1\textwidth]{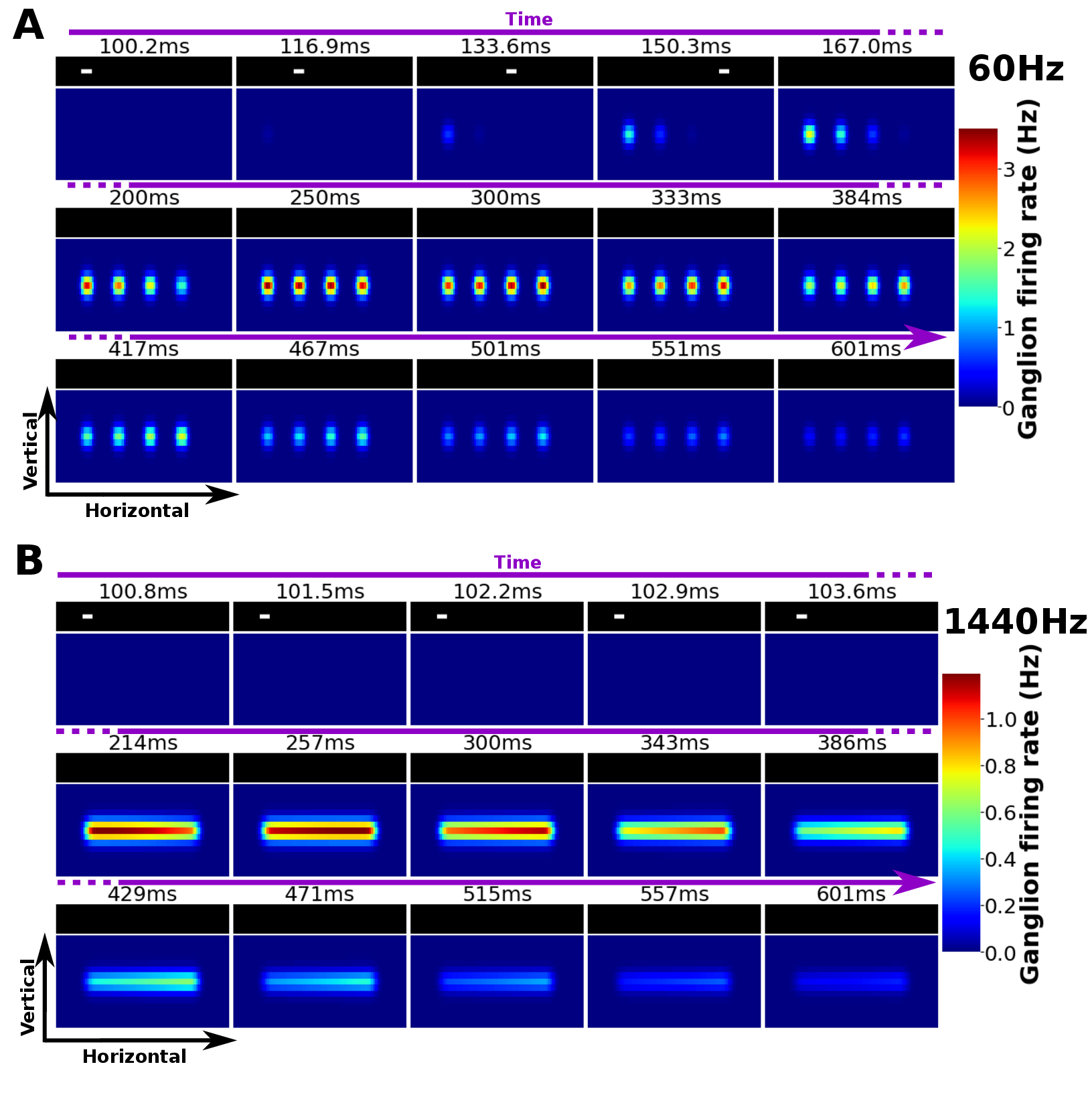}
\caption{
\textbf{The retinal activity generated by a fast movement is strongly influenced by the frame rate.} For each figure, the time axis is represented by the top purple arrow.  We display time snapshots indicated below the time axis. Black rectangles show the successive positions of the bar. Color images show the RGCs activity (color map) at times indicated on the top. The first five images are the first five frames of our video, in order to show the bar positions (in
the fifth frame it has already left the visual field). The other ten images are frames separated by a longer time interval to illustrate the integration by RGCs.
\textbf{A)} \textbf{Frame rate of $60$ Hz.} 
\textbf{B)}\textbf{Frame rate of $1440$ Hz.} 
}
\label{Fig:FrameRate_effect_high_speed_retina}
\end{figure}
\begin{figure}[!ht]
\centering
\includegraphics[width=1\textwidth]{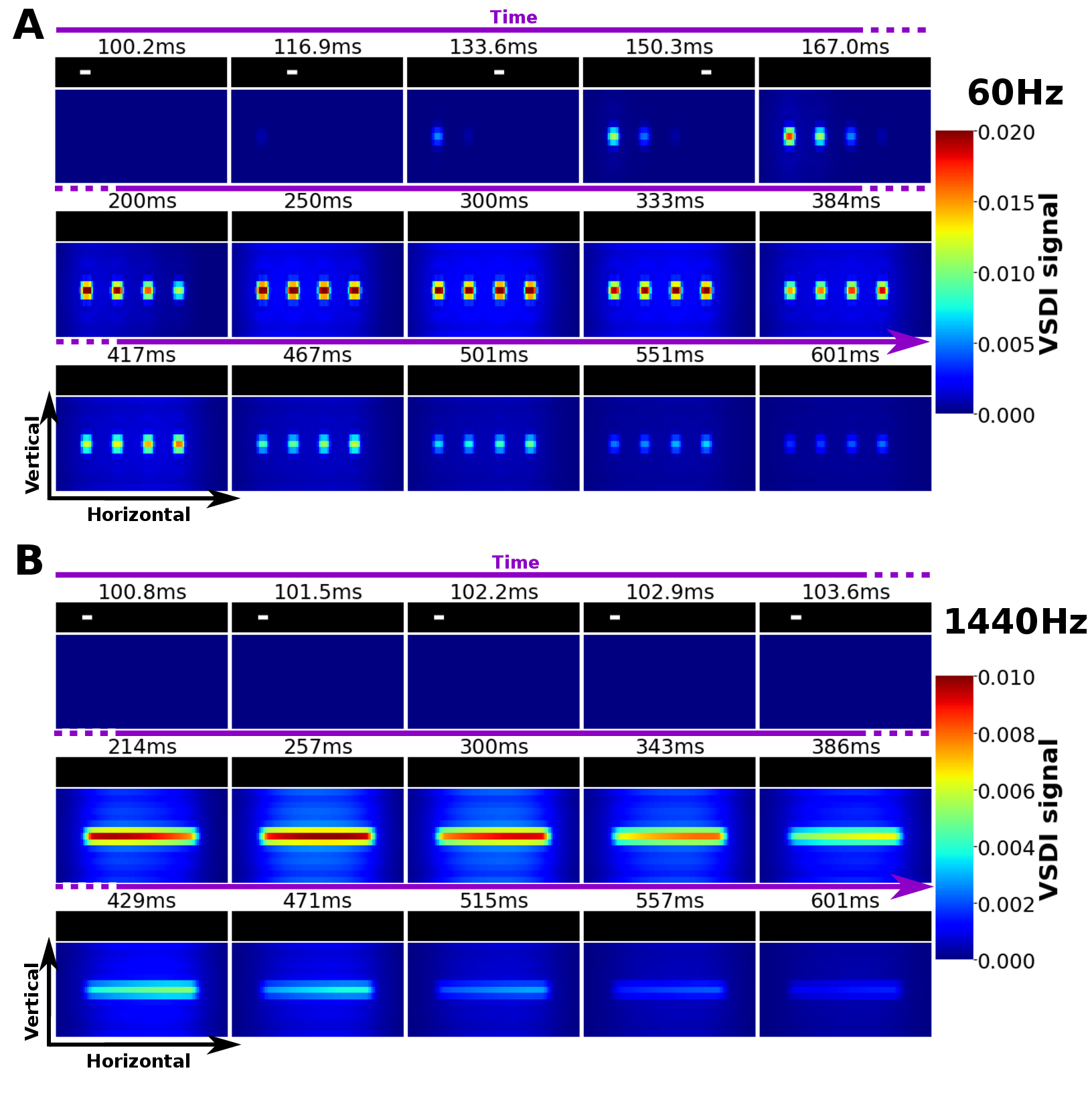}
\caption{
\textbf{The cortical activity generated by a fast movement is strongly influenced by the frame rate.} Same representation as \ref{Fig:FrameRate_effect_high_speed_retina} but here the color maps correspond to cortical activity (VSDI).
}
\label{Fig:FrameRate_effect_high_speed_cortex}
\end{figure}

%
The movie of the bar was designed so that the cortical projection of the bar has a size of $1.08\degree \times 0.45\degree$ along a trajectory of $12\degree$. To remove boundaries effects, the bar appears at the position $3.25\degree$, at time $100$ ms, and then disappears at the position $15.25\degree$ 
$60$ ms later. The piece of cortex have a size of $18.45\degree \times 3.15\degree$, see \cite{emonet:24} for detail.
%
The time interval between two images, $\delta_{t}$, the inverse of the refresh rate, takes the value $0.0167$ s ($60$ Hz) or $0.0007$ s ($1440$ Hz). 
At $1440$ Hz, during the time necessary for the bar to cross the visual field,
the positions taken by the bars are more numerous and closer ($0.14\degree$ per frame)
than at $60$ Hz ($3.34\degree$ per frame, corresponding to only $4$ bar positions).
Figure \ref{Fig:FrameRate_effect_high_speed_retina} compares the temporal sequences of the retinal response (RGC firing rate) obtained with the two refresh rates. For high frame rate, the response to the bar seems continuous. In contrast, for the low frame rate, the retinal response observed looks discontinuous.
The same observations hold for the cortical response (Fig. \ref{Fig:FrameRate_effect_high_speed_cortex}). 

The discontinuity of response at $60$ Hz is of course induced by the discontinuity of the stimulus, where, at each frame the bar makes a huge spatial leap.
However, if one increased continuously the frame rate, there should be a switch in the perception, from discontinuous to continuous.
Which parameters, or as we show now, which \textit{combination of parameters}, in the stimulus and in the retino-cortical model, determine this switch ?
To answer this question we provide a simple mathematical argument.


\subsection{The phenomenological parameters that fix the perception} \label{Sec:params}



\ssSec{Analytic development}{Analytic}

We consider a moving bar, with speed $v$ along the horizontal axis, infinite height, width $2b$. The movie of this bar is projected to the retina with an overhead projector having a refresh rate $\delta_t$.
The corresponding stimulus reads mathematically:
\begin{equation}\label{eq:StimRefresh}
\cS(x,y,t) =  \sum_{n=-\infty}^{+\infty}  \Pi_b\pare{x-n \,v \delta_t} \, 
\chi\pare{t \in [n\,\delta_t, (n+1)\,\delta_t [},
\end{equation}
where $\Pi_b\pare{u}=\left\{ 
\begin{array}{lll}
1,  \quad &\mbox{if} \, -b \leq u \leq b\\
0, & \mbox{otherwise},
\end{array}
\right.$
is the analytic form of the bar. Here $\chi$ is the indicator function of events ($\chi\pare{A}=1$ if $A$ is true, $\chi\pare{A}=0$ if $A$ is false), formalising the notion of frame duration: time is continuous but the image in frame $n$ stays still from time $n \delta_t$ to time $(n+1)\,\delta_t $ where it is replaced by a new image. For simplicity, we have considered 
an infinite motion in the past and in the future.

In the model processing, from the photo-receptors layer to RGCs then to V1, the response corresponds to a cascade of convolutions, rectified by non linearities. Here, we consider only a linear cascade. Thus, the response of a RGC, or of a cortical column, is a convolution of the stimulus with a spatio-temporal kernel corresponding to the receptive field of this RGC or cortical column. For simplicity, we assume a separable kernel of the form:
\begin{equation}\label{eq:K}
\cK(x,y,t)=\cK_S(x,y) \, \cK_T(t),
\end{equation}
where $\cK_S$ is a Gaussian that mimics the integration of the center part of the RGC receptive field with a characteristic size $\sigma_c$. Thus, we do not consider surround effects in the next computations, but this could be done similarly (we would have a difference of terms coming from the integration on a difference of Gaussian).
The temporal kernel, $\cK_T$, has a characteristic time $\tau$. Here, we consider an exponential kernel\\ $\cK_T(t)=\frac{1}{\tau} \,e^{-\frac{t}{\tau}} \, H(t)$ where $H(t)$ is the Heaviside function.

The spatio-temporal convolution of the stimulus \eqref{eq:StimRefresh} with this kernel can be analytically computed. This is:
\begin{equation}\label{eq:Vdrive}
\cR_i(t) \,=\,
\int_{x=-\infty}^{+\infty} \,\int_{y=-\infty}^{+\infty} \,\int_{s=-\infty}^{t} \,  \cK(x-x_i,y-y_i,t-s) \, \cS(x,y,s) dx \, dy \, ds 
\,=\,
\sum_{n=-\infty}^{n_t} \, \cG(n,t)\, \cP_i(n \delta_t)
\end{equation}
where $n_t=\bra{\frac{t}{\delta_t}}$ is the integer part of $\frac{t}{\delta_t}$,
and where:
\begin{equation}\label{eq:CGT}
 \cG(n,t)=
    \left\{
\begin{array}{lll}
     &  0, \quad \mbox{if} \quad t<n\,\delta_t;\\
     & 1-e^{-\frac{t-n\,\delta_t}{\tau}}, \quad \mbox{if} \quad n\,\delta_t \leq t < (n+1)\,\delta_t;\\
     & e^{-\frac{t-(n+1)\delta_t}{\tau}}\bra{1-e^{-\frac{\delta_t}{\tau}}}, \quad \mbox{if} \quad t \geq (n+1)\,\delta_t.
\end{array}
\right.
\end{equation}
is the time integral of the kernel $\cK_T$ during the frame $n$. This corresponds to a low pass filter integration, with characteristic time $\tau$. This function lets appear  ratio:
\begin{equation}\label{eq:r}
r=\frac{\delta_t}{\tau},
\end{equation}
between the time length of a frame and the characteristic time of integration $\tau$.

In eq. \eqref{eq:Vdrive}, the term:
\begin{equation}\label{eq:Pi}
\cP_i(t)=\frac{1}{2} \, \pare{Erf\pare{D^+_i(t)} -Erf\pare{D^-_i(t)}},
\end{equation}
comes from the spatial integration of the bar, for a cell located at coordinate $x_i$ (the integration along the $y$ axis does not play a role here as we consider an infinite bar). We have set:
\begin{equation}\label{eq:Di}
D^\pm_i(s)=\frac{\pm b -x_i + v \, s}{\sigma_c},
\end{equation}
the algebraic distance of the leading ($+ b$), trailing ($-b$) edge of the bar, to the center of the RF at time $t$, measured in units of spatial RF ($\sigma_c$). We have also set $\Erf(x)=\erf(x/\sqrt{2})$

The function $\cP_i(t)$ corresponds to a moving pulse, propagating at speed $v$ whose shape depends the dimensionless ratio:
\begin{equation} \label{eq:kappa}
    \kappa=\frac{1}{3}\frac{b}{\sigma_c},
\end{equation}
where the coefficient $\frac{1}{3}$ has been added so that the value $\kappa=1$ corresponds to transition in the shape of $\cP_i(t)$. Indeed, $\kappa$ compares the size of the bar, $2b$, to the size of the RF, that is the size of the region whose spatial integration is essentially $1$ (corresponding to a spatial extent $\sim 2\times 3 \sigma_c$) . If $\kappa \ll 1$ the bar is very thin compared to the RF and $\cP_i(t)$ is a Gaussian like pulse. In contrast, when $\kappa \gg 1$, the bar is very large. When it enters the RF there is a fast rise of activity (corresponding to the erf function with a + sign in \eqref{eq:Pi}) then a plateau when the bar fills the full RF. Then there is a fast decay when the bar gets outs (erf function with the - sign). In this regime, the time plateau has a width $\frac{2b}{v}$, the time required for the bar to fully cross a given position. Finally, the intermediate value $\kappa \sim 1$ corresponds to a switch between the Gaussian pulse shape and the flat pulse shape.

The pulse $\cP_i$ is a difference of $\erf$ functions. Considering that $\Erf\pare{\frac{u}{\sigma_c}} \sim 0$
if $u<-3 \sigma_c$ and $\Erf\pare{\frac{u}{\sigma_c}} \sim 1$ if $u >3 \sigma_c$, $\cP_i(n \delta_t)$ is non zero
if 
%
%
$n \in \bra{\frac{-b+x_i-3\sigma_c}{v \delta_t},\frac{b+x_i+3\sigma_c}{v \delta_t}}$. This gives the approximate number of frames that the cell located at $x_i$ is going to "see", about:
\begin{equation}\label{eq:Nperc}
N_{perc}\,=\,\bra{2\frac{b+3\sigma_c}{v\delta_t}}\,=\, \bra{\frac{r}{\rho} \pare{\kappa+1}}
\end{equation}
frames, where we have introduced:
\begin{equation}\label{eq:rho}
\rho=\frac{v \tau}{6 \sigma_c}=\frac{\tau}{\tau_{cross}},
\end{equation}
the distance covered by the bar during the characteristic time of the cell's response, compared to the size of the RF. Equivalently, this the ratio between $\tau$ and $\tau_{cross}=\frac{6\sigma_c}{v}$ the time taken by the bar to cross the RF.

Equation \eqref{eq:Vdrive} tells us that the response of the cell located at $x_i$ is the discrete superimposition 
of moving pulses, $\cP_i$, shifted by the discrete time jump $n \delta_t$ and weighted by \eqref{eq:CGT}, the time integration of the receptive field over the time where the image stays still.

There are several cases, depending on the number of perceived frames \eqref{eq:Nperc}, constrained by the $3$ dimensionless parameters: $\kappa$, which controls the shape of the spatial response; $r$, which controls the ratio between the frame rate and the characteristic integration time; and $\rho$, which controls the ratio between the characteristic integration time and the time taken by the bar to cross the RF.

\begin{enumerate}
   \item \textbf{No response.} If $N_{perc}=0$, ($v \delta_t \gg 2\pare{b+3\sigma_c}$), the bar goes so fast that the cell has no time to see even one frame.
    \item \textbf{Flashed image.} If $N_{perc}=1$, ($v \delta_t \sim 2\pare{b+3\sigma_c}$), the cell perceives only one frame, i.e. it sees a unique image, flashed during the time $\delta_t$.
    \item \textbf{Saccadic refreshing.} When $N_{perc}=k>1$  the cell receives $k$ images, separated by a time interval $\delta_t$. If $\delta_t  \gg \tau$, i.e. $r \gg 1$ the time integration follows the jumps induced by the spatial shifts of size $v\delta_t$ in the bar
motion. The cell perceives a discontinuous, jerky, scene.  The movement is perceived as a succession of static image (flashes). 
This is what we observe in Fig. \ref{Fig:FrameRate_effect_high_speed_retina} A, \ref{Fig:FrameRate_effect_high_speed_cortex} A.
\item \textbf{Continuous perception.} 
In contrast, if  $\delta_t  \ll \tau$, i.e. $r \ll 1$ the low pass integration somewhat smooths the perceived motion.  The movement is perceived as continuous.
This is what we observe in Fig. \ref{Fig:FrameRate_effect_high_speed_retina} B, \ref{Fig:FrameRate_effect_high_speed_cortex} B.
A particular case of this arises when $\delta_t \to 0$, whereas $v$ is fixed. Here, $N_{perc} \to +\infty$ and the limit corresponds to a continuous integration.
\item \textbf{Intermediate case.} When  $r\sim1$, there is an intermediate regime where the response is a sum of low pass filters integration and where the cell has just enough time to react fully before the frame changes. The movement will therefore be perceived, but more or less jerky. Here, $r$ is not sufficient to determine what the perception will be.  
\end{enumerate}

This analysis tells us that, if the perception is controlled by $r$ it is also dependent on the bar speed (via $\rho$) and on the bar width (via $\kappa$). Thus,  $r$ is not enough to characterize the perception. For example, in the case of Fig. \ref{Fig:FrameRate_effect_high_speed_cortex},
we have  $\tau \sim 0.1$ s corresponding to the largest characteristic time (RGCs) in the cascade of integration. A frame rate of $60$ Hz ($\delta_{t}=0.0167$) gives $r=0.167 < 1$ while the observed response at the cortical level is shown in Fig. \ref{Fig:FrameRate_effect_high_speed_cortex}A, a succession of static images. This corresponds to the  intermediate case (item 5 above), where $r$ is not enough to determine what the perception will be. In contrast, a frame rate of $1440$ Hz ($\delta_{t}=0.007$) gives a $r=0.07 \ll 1$ and a continuous perception (Fig. \ref{Fig:FrameRate_effect_high_speed_cortex}B).

\section{Discussion}\label{Sec:Discussion}

This communication provides an heuristic and quantitative argument, supported by numerical simulations, suggesting that the refresh rate of overhead projectors can affect the visual perception of fast moving objects. Although, our simulations are limited to the early visual system (retina-V1) the effect we observe ought to be propagated at higher levels of visual processing, in agreement with psycho-physics experiments made by M. Wexler. Our analysis shows that the perception is not only dependent on the refresh rate, it depends on physiological characteristics, the size of the receptive field and the characteristic time of response, and the stimulus itself (here the bar width and bar speed). In general, the perception is that of a sequence of flashed images which can appear continuous or discontinuous depending on these parameters, and summarized by the $3$ dimensionless parameters \eqref{eq:r}, \eqref{eq:kappa}, \eqref{eq:rho}.

We would like now to discuss several  caveats and potential extensions of this approach. The main criticism is of course the simplicity of the model, and even more, of the mathematical argumentation which neglects a lot of important aspects. First, it is evident that the retina and the visual cortex are quite more complex than the model we used, and, when considering the response of a moving object such effects as the nonlinearities at the level of photo-receptors with a high dependence on contrast ought to be considered. Indeed, the contrast of the moving object might have a important effect on the perception, in addition to the refresh rate. Also, 
the retina is a network involving a large range of time and space scales \cite{kartsaki-hilgen-etal:24} whereas our mathematical argumentation considers only one space and time scale. It could also be relevant to consider the effect of surround on the response. In this case, from the same computation done in eq. \eqref{eq:Vdrive} we would expect that a negative surround would enhance the effect of slow refresh, while smoothing more the high resfresh response. Finally, our study does not differentiate between retinal cell types (ON-OFF, magno-parvo) and  the subsequent pathways. We may expect substantial differences of response between these cells type. It would be interesting to know whether our quantitative characterisation \eqref{eq:Nperc} would help discriminating the variability in the responses of these cells. Clearly, experimental investigations, such as the one currently performed on the retina in O. Marre's lab (private communication) are absolutely inescapable. 

\textbf{Acknowledgements}
This work is funded by the ANR Shooting Star-15755. We are grateful to Mark Wexler, Frédéric Chavane, Alain Destexhe, and more especially to Olivier Marre and Awen Louboutin for insightful discussions.

 \bibliographystyle{apalike}
\bibliography{biblio.bib} 

\begin{thebibliography}{}

\bibitem[Berry et~al., 1999]{berry-brivanlou-etal:99}
Berry, M., Brivanlou, I., Jordan, T., and Meister, M. (1999).
\newblock Anticipation of moving stimuli by the retina.
\newblock {\em Nature}, 398(6725):334---338.

\bibitem[Campbell and Wurtz, 1978]{campbell-wurtz:78}
Campbell, F.~W. and Wurtz, R.~H. (1978).
\newblock Saccadic omission: {Why} we do not see a grey-out during a saccadic
  eye movement.
\newblock {\em Vision Research}, 18(10):1297--1303.

\bibitem[Cessac, 2022]{cessac:22}
Cessac, B. (2022).
\newblock Retinal processing: Insights from mathematical modelling.
\newblock {\em Journal of Imaging}, 8(1).

\bibitem[Chemla, 2010]{chemla:10}
Chemla, S. (2010).
\newblock {\em A biophysical cortical column model for optical signal
  analysis}.
\newblock PhD thesis, School of Information and Communication Sciences.

\bibitem[Chen et~al., 2013]{chen-marre-etal:13}
Chen, E.~Y., Marre, O., Fisher, C., Schwartz, G., Levy, J., da~Silviera, R.~A.,
  and Berry, M. (2013).
\newblock Alert response to motion onset in the retina.
\newblock {\em Journal of Neuroscience}, 33(1):120--132.

\bibitem[Di~Volo et~al., 2019]{di-volo-romagnoni-etal:19}
Di~Volo, M., Romagnoni, A., Capone, C., and Destexhe, A. (2019).
\newblock {Biologically realistic mean-field models of conductance-based
  networks of spiking neurons with adaptation}.
\newblock {\em {Neural Computation}}, 31(4):653--680.

\bibitem[Duyck et~al., 2018]{duyck-wexler-etal:18}
Duyck, M., Wexler, M., Castet, E., and Collins, T. (2018).
\newblock Motion {Masking} by {Stationary} {Objects}: {A} {Study} of
  {Simulated} {Saccades}.
\newblock {\em i-Perception}, 9(3).

\bibitem[Emonet, 2024]{emonet:24}
Emonet, J. (2024).
\newblock {\em A retino-cortical model to study movement-generated waves in the
  visual system}.
\newblock Computer science, Nice C{\^o}te d'Azur, Valbonne.

\bibitem[Emonet et~al., 2025]{emonet-souihel-etal:24}
Emonet, J., Souihel, S., Di~Volo, M., Destexhe, A., Chavane, F., and Cessac, B.
  (2025).
\newblock A chimera model for motion anticipation in the retina and the primary
  visual cortex.

\bibitem[Kartsaki et~al., 2024]{kartsaki-hilgen-etal:24}
Kartsaki, E., Hilgen, G., Sernagor, E., and Cessac, B. (2024).
\newblock {How does the inner retinal network shape the ganglion cells
  receptive field : a computational study}.
\newblock {\em Neural Computation}, 36(6):1041--1083.

\bibitem[Souihel and Cessac, 2021]{souihel-cessac:21}
Souihel, S. and Cessac, B. (2021).
\newblock On the potential role of lateral connectivity in retinal
  anticipation.
\newblock {\em J. Math. Neurosc.}, 11(3).

\bibitem[Westheimer, 1954]{westheimer:54}
Westheimer, G. (1954).
\newblock Mechanism of saccadic eye movements.
\newblock {\em A.M.A. Archives of Ophthalmology}, 52:710--723.

\bibitem[Zerlaut et~al., 2018]{zerlaut-chemla-etal:18}
Zerlaut, Y., Chemla, S., Chavane, F., and Destexhe, A. (2018).
\newblock Modeling mesoscopic cortical dynamics using a mean-field model of
  conductance-based networks of adaptive exponential integrate-and-fire
  neurons.
\newblock {\em Journal of Computational Neuroscience}.

\end{thebibliography}

\end{document}